\documentclass[twocolumn,amsmath,amssymb]{revtex4}
\usepackage{graphicx}
\usepackage{dcolumn}
\usepackage{amsmath}
\usepackage{bm}
\DeclareGraphicsExtensions{.pdf,.png,.jpg}
\usepackage{epstopdf}
 
\begin{document}
 
\title{Superconductivity on the verge of electronic topological transition in Fe based superconductors}
\author{Haranath Ghosh, Smritijit Sen}
\affiliation{Indus Synchrotrons Utilization Division, Raja Ramanna Centre for Advanced Technology, 
Indore -452013, India. \\  and Homi Bhabha National Institute, Anushakti
Nagar, Mumbai 400094, India.} 
\begin{abstract}
Doping as well as temperature driven Lifshitz transitions are 
found from first principles simulations in a variety of Fe based superconductors that are
 consistent with experimental findings. In all the studied compounds the Lifshitz transitions 
 are consistently found to occur at a doping concentration where superconductivity 
 is highest and magnetism disappears. Systematically, the Lifshitz transition occurs in the 
 electron Fermi surfaces for hole doping, whereas in hole Fermi surfaces for electron doping 
 as well as iso-electronic doping. Temperature driven Lifshitz transition is found to occur 
 in the iso-electronic Ru-doped BaFe$_2$As$_2$ compounds. Fermi surface areas are found to carry 
 sensitivity of topological modifications more acutely than the band structures and can be used as 
 a better experimental probe to identify electronic topological transition. 
\vspace{1pc}
\end{abstract}
\maketitle


Electronic topological transition of the Fermi surface with no broken symmetries, known as Lifshitz 
transition plays a significant role in numerous branches of condensed mater physics. For example, 
the collapse of the normal state pseudogap at a Lifshitz transition in Bi$_2$Sr$_2$CaCu$_2$O$_{8+\delta}$ 
cuprate superconductor \cite{Benhabib}, Lifshitz transition in underdoped cuprates \cite{Norman}, 
Dirac semi metal \cite{Su-Yang Xu}, relativistic insulator NaOsO$_3$ \cite{Bongjae}, discontinuous 
Lifshitz transition in  Na$_x$CoO$_2$ \cite{Okamoto}, Zeeman driven Lifshitz transition in 
YbRh$_2$Si$_2$ \cite{Hackl}, in two-dimensional Hubbard model \cite{Chen}, bilayer graphene \cite{Lemonik}, 
Quantum Hall Liquids \cite{Varlet}, to mention a few. 
Lifshitz transition is recently considered as a quantum phase transition in strongly 
correlated electron systems also plays a remarkable role in Fe-based high T$_c$ superconductors 
owing to its special `Fermiology' and multi-band nature \cite{LiuA,Khan}. 
The phase diagrams of Fe-based superconductors (SCs) consist of a number of exhilarating features 
apart from its novel superconducting phase \cite{Nandi,Avci,Ni,Thaler,Kasahara}. 
Various phases are very much sensitive to external parameters like impurity, doping, 
temperature and pressure. 
On the other hand, Lifshitz transitions are experimentally observed with respect to some of the 
above mentioned external parameters \cite{Liu,Xu,LiuB}. 
Therefore, topological transitions, magnetism and superconductivity which occurs
in a variety of materials in condensed matter physics is a
subject of general interest.

 The Fe-based high T$_c$ SCs being a multiband system has multiple Fermi surfaces (FS) (two electron like and 
three hole like), any of the FSs may collapse (become smaller ones) either because, some of the bands 
that crosses the Fermi level (FL) may move away (above or below) from the FL or due to topological modifications 
({\it e.g}, electron like band transforming to hole like or vice versa) under external perturbation.   
Lifshitz transition (LT) or electronic topological transition (ETT), where FSs/electronic bands alter 
topology, have been usually investigated as zero temperature phenomena arising due to impurity, doping 
or pressure etc. However, temperature dependent LT/ETT is observed recently in WTe$_2$ \cite{Kaminski}. 
The consequences of these LT/ETT are innumerable. It can lead to reduced interband scattering (affecting mechanism of SC), nesting of FS (affecting magnetism), anomalies or singularities in the density of states at FL and in general anomalies in the kinematics, 
dynamics and thermodynamics of electrons, which would affect various physical 
properties \cite{Lifshitz}. LT/ETT are also observed in Fe-based SCs induced by pressure, 
impurity and doping  \cite{Quader,Sn,Khan,Liu,Wang}. 
In particular, for BaFe$_2$As$_2$ (Ba122) system, LT is predicted theoretically due to small but 
unintentional Sn impurity in contrast to other  compounds like SrFe$_2$As$_2$ (Sr122) and CaFe$_2$As$_2$ (Ca122) \cite{Sn}. 
In hole doped Ba122, LT is  found in the heavily doped regime as 
evident from the theoretical work of Khan {\it et al}., \cite{Khan}. On the other hand, 
LT in electron (Co in place of Fe) doped Ba122 is observed experimentally \cite{Kaminski}. 
But the temperature induced LT in general is rare;
we have predicted temperature dependent LT in Ru doped Ba122 system. 
In this letter, we show with explicit demonstrations in a series of Fe-based SCs  
that the LT / ETT occur at a doping concentration where SC is maximum and magnetism/structural transition vanishes.

Determining LT/ETT experimentally is challenging specially in systems that constitute multi-orbital-derived FS like Fe-based SCs; we show that
the FS area (FSA) carries sensitivity of topological modifications more acutely than the band structures (BS) and this can be used 
as a better experimental tool to identify ETT/LT.
As a result of LT or ETT the FSA of a particular FS gets largely affected and hence a systematic study 
of FSA as a function of doping is most desirable.  A detailed study on the variations of FSA, {\it e.g.}, variations of (i) areas of each individual FSs, (ii) sum total areas of all the electron FSs,  (iii) sum total areas of all the hole 
FSs, (iv) sum total areas of all the five FSs, (v) difference of all hole and all electron FS areas as a 
function of doping is a rare wealth of information that can be verified by the de Haas-van Alphen and allied effects (i.e, Shubnikov-de Haas effect) are presented in this letter. All the above FS areas with doping show distinctly different behaviors below and above the ETT. 
Clear deviation in the variations of FSA with doping marks the occurrence of
topological transition (TT). This can therefore lead to a step forward progress in the experimental detection of
LT/ETT in general. It is worthwhile mentioning here that in the overdoped regime of high-T$_c$ cuprate SCs,
 ARPES and quantum oscillation studies revealed large FS but not only its shape and size change with doping 
 but also breaks up into Fermi arcs \cite{Norman1} which turned out to be closed hole 
 Fermi pocket \cite{hbyang,hbyang1,LeBoeuf,Sebastian}. 
 All these points to a question, are high T$_c$ SCs topological or whether the high  T$_c$ is 
 restricted by TT ?!

  \begin{figure*}[t]
  \flushleft \includegraphics [width=\textwidth,height=5cm]{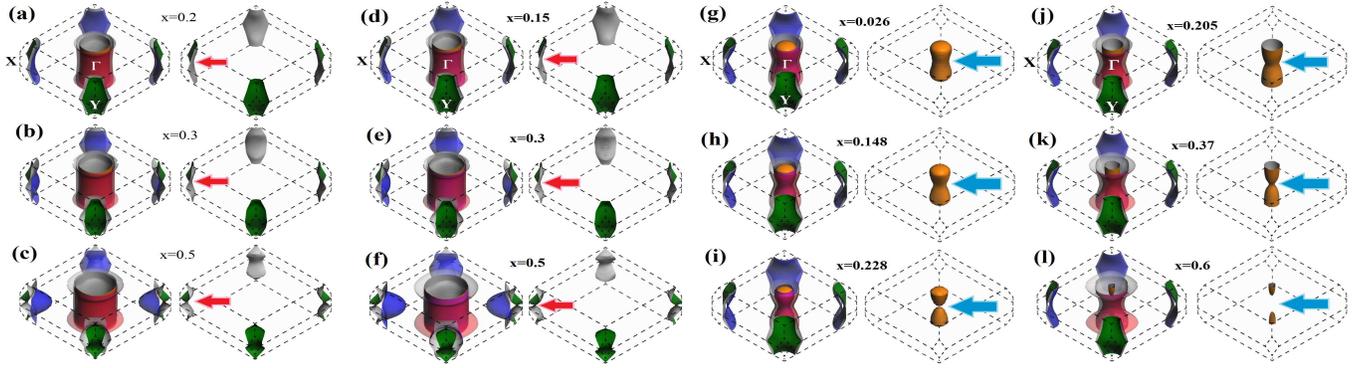}
    \caption{(Colour online) Calculated FSs in four pair of columns for (i) Ba$_{1-x}$K$_x$Fe$_2$As$_2$  (a \----c); (ii) Ba$_{1-x}$Na$_x$Fe$_2$As$_2$ (d \---- f) in the low temperature orthorhombic
        phase; (iii) BaFe$_{2-x}$Co$_x$As$_2$ (g\----i) in the high temperature 
          tetragonal phase and (iv) BaFe$_2$(As$_{1-x}$P$_x$)$_2$ (j\----l)  in the low temperature orthorhombic phase for various dopings as indicated in the figure. In the left columns of each pair, all the five FSs around $\Gamma$ (hole like), 
     X and Y (electron like) points are shown using different colours for each compounds whereas in the right columns of the same pair, topologically modified FSs are shown separately.}
   \label{KFS}
  \end{figure*}
   \begin{figure}[ht]
     \centering
 \includegraphics [height=3.0cm,width=0.45\textwidth]{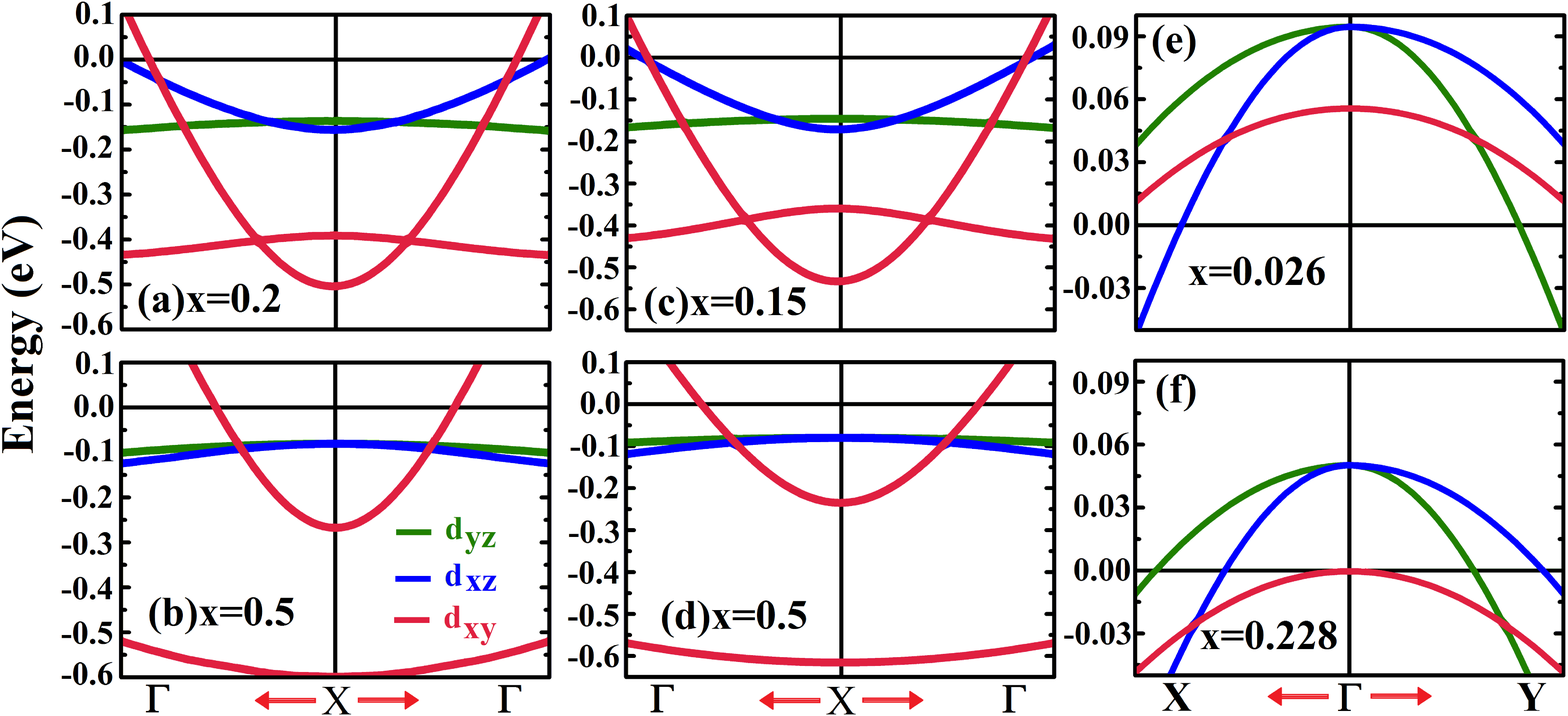}
     \includegraphics [height=3.0cm,width=0.45\textwidth]{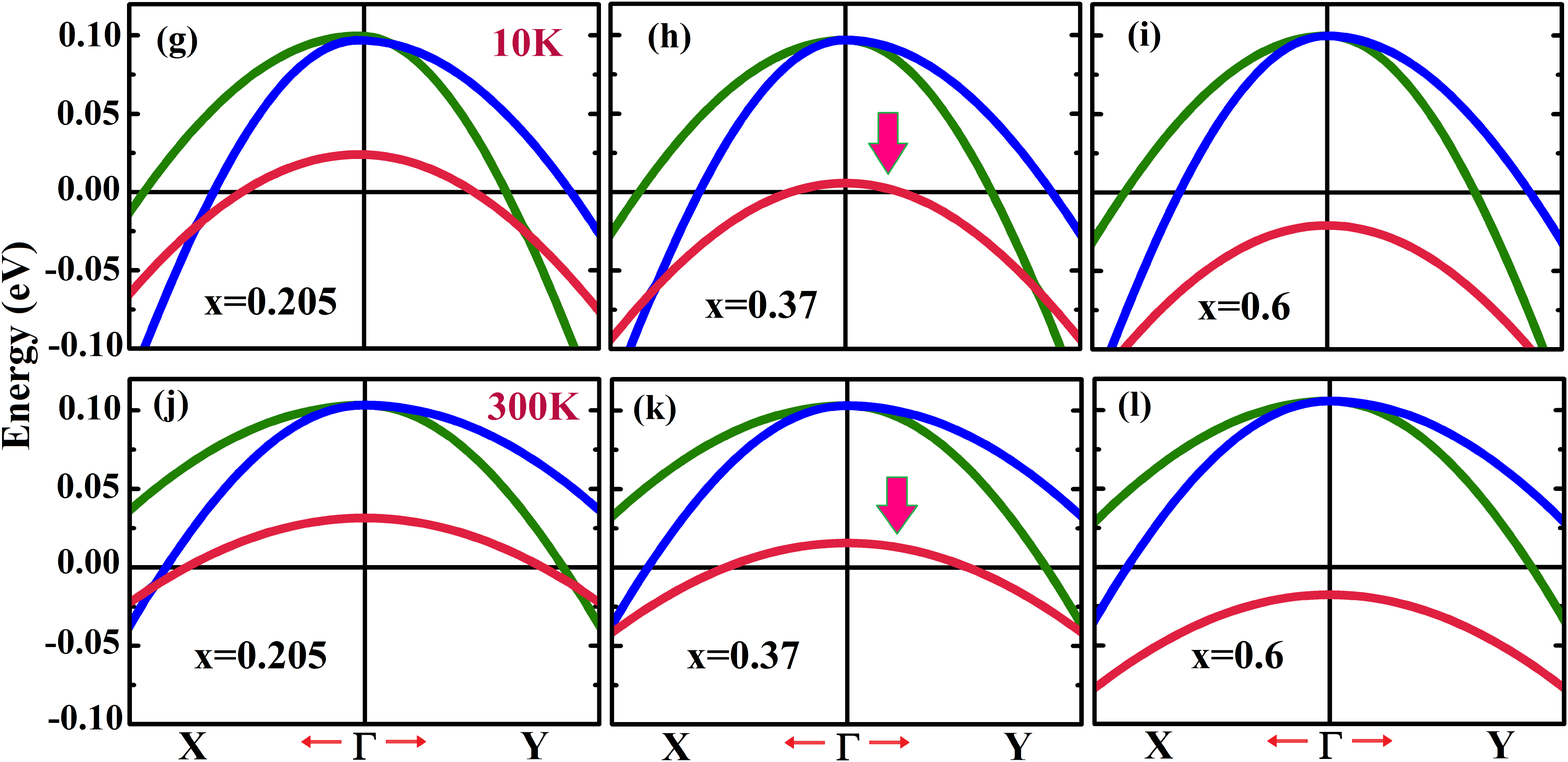}
     \caption{(Colour online) Zoomed orbital resolved electronic BSs around the high symmetry points where TTs are found in Figs.\ref{KFS} for (i) Ba$_{1-x}$K$_x$Fe$_2$As$_2$  (a \--b); (ii) Ba$_{1-x}$Na$_x$Fe$_2$As$_2$ (c \-- d) ; (iii) BaFe$_{2-x}$Co$_x$As$_2$ (e\--f) and (iv) BaFe$_2$(As$_{1-x}$P$_x$)$_2$ (g\----l). Topological modifications in the d$_{xz}$, d$_{xy}$ bands from electron to hole like dispersion is observed as doping is increased (cf. (a,b) and (c,d)). LT in the d$_{xy}$ bands of BaFe$_{2-x}$Co$_x$As$_2$ [Figs.(e\--f)] and BaFe$_2$(As$_{1-x}$P$_x$)$_2$ [Figs.(g\----l)] are worth observing.
         }
     \label{BS-e-h-iso}
     \end{figure}
First principles density functional theories (DFT) can produce correct solutions of the many electron Schr\"{o}dinger equation if exact electronic density is  used as input. Various modern X-ray diffraction techniques {\it e.g}., using Synchrotrons 
radiation source etc. that determines crystallographic information at different external perturbations are essentially 
result of diffraction from various atomic charge densities (Bragg's diffraction). Considering experimentally determined 
structural parameters at different temperatures (doping)  as input thus in turn provides temperature (doping) dependent 
exact densities in our first principles calculation. These input structural parameters are kept fixed through out the calculation 
for a fixed temperature (doping). The main effect on the electronic structure from finite temperature is the underlying crystal structure, 
and the average crystal structure at finite temperature can usually be reliably determined from diffraction experiment at a given temperature. Our first principles calculations are performed by implementing ultrasoft pseudopotential with plane wave basis set based on DFT \cite{CASTEP}. Electronic 
exchange correlation is treated within the generalised gradient approximation (GGA) using 
Perdew-Burke-Ernzerhof (PBE) functional \cite{PBE}. 
Electronic structures (BS, FS {\it etc.}),
calculated using optimized lattice parameters ($a$, $b$, $c$ and $z_{As}$) do not resemble with that of the 
experimentally measured one \cite{pla,zAs}. This insist us to employ experimental 
lattice parameters {\it i.e.,} $a$, $b$, $c$ and $z_{As}$ 
\cite{Avci,thesis,Acta,Sefat,Ni} as a function of doping and temperature both in the low temperature
orthorhombic as well as in the high temperature tetragonal phases
as inputs of our first principles calculations. 
In order to dope, we use Virtual crystal approximation (VCA) \cite{Bellaiche}.
Non-spin-polarized and spin polarised single point energy calculations 
are performed for tetragonal phase with space group symmetry I4/mmm (No. 139) and 
orthorhombic phase with space group symmetry Fmmm (No. 69) respectively 
with energy cut off 500 eV and higher as well as
self-consistent field (SCF) tolerance as $10^{-7}$ eV/atom. 
Brillouin zone is sampled in the k-space within Monkhorst–Pack scheme
and grid size for SCF calculation is chosen as per requirement of the calculation 
for different systems. For simulating FS, grid size of SCF calculation is chosen as
$26\times 26\times 31$.

\begin{figure}[ht]
  \centering
  \includegraphics [height=5.0cm,width=0.45\textwidth]{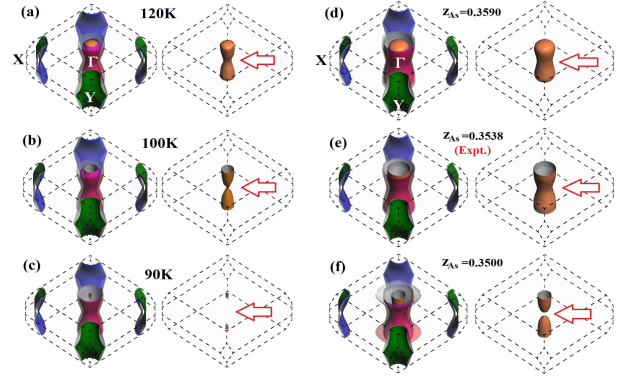}
  \caption{(Colour online) Calculated FS of BaFe$_{2-x}$Ru$_x$As$_2$ below structural transition 
  (pair of left columns) and
  that of undoped BaFe$_2$As$_2$ for different z$_{As}$ (pair of right columns). From each pair of columns, in the left column all the five FSs around $\Gamma$ (hole like), X and Y (electron like) points are displayed using different colours while in the right column topologically modified inner hole FSs are indicated by arrows.}
  \label{RuFS}
 \end{figure}

We investigate the role of various kinds of dopings ({\it e.g}, electron, hole or iso-electronic)  in
ETT / LT of various Fe-based superconductors [Ba$_{1-x}$M$_x$Fe$_2$As$_2$ (M= K, Na) (see Figs. \ref{KFS} (a\---f)), BaFe$_{2-x}$Co$_x$As$_2$ (see Figs. \ref{KFS} (g\---i)),  BaFe$_2$As$_{2-x}$P$_x$ (see Figs. \ref{KFS} (j\---l))]. 
In Fig.\ref{KFS}(a\---c) leftmost pair of columns, we depict the FSs of K doped Ba122 systems. 
It is very clear from Fig.\ref{KFS} that there are five Fermi pockets, 
two around X and Y points (electron like) and three around $\Gamma$ point 
(hole like). 
We find that one of the electron like FSs around X/Y point 
transform into a different topology ({\it cf}. right column Fig. \ref{KFS}(c)). 
We display orbital projected BS 
of Ba$_{1-x}$K$_x$Fe$_2$As$_2$ around X point separately in Fig.\ref{BS-e-h-iso}(a,b) for $x=0.2$ and $x=0.5$ respectively. Same colour codes are used for all orbital resolved BSs throughout. 
From Fig.\ref{BS-e-h-iso} (a,b) it is quite clear that the  electron like nature of d$_{xz}$ band around X (Y) points transform into hole like band; one of the hole like d$_{xy}$ bands also gets modified to an electron like band as a result of K-doping. These topological transformations of FSs/BSs are identified as ETT or LT.
 \begin{figure}[ht]
  \centering
  \includegraphics [height=1.75cm,width=0.42\textwidth]{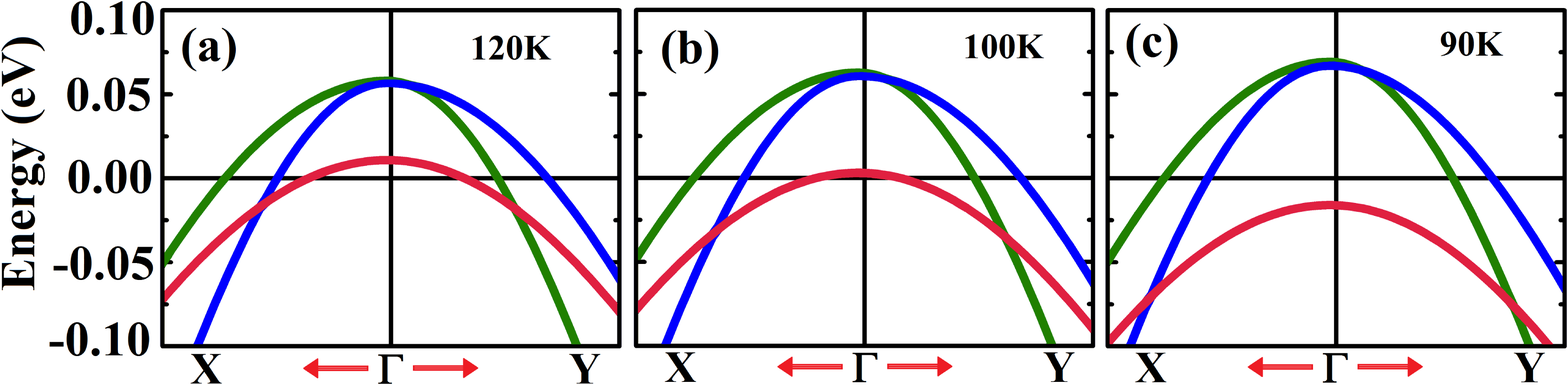}
  \includegraphics [height=1.75cm,width=0.43\textwidth]{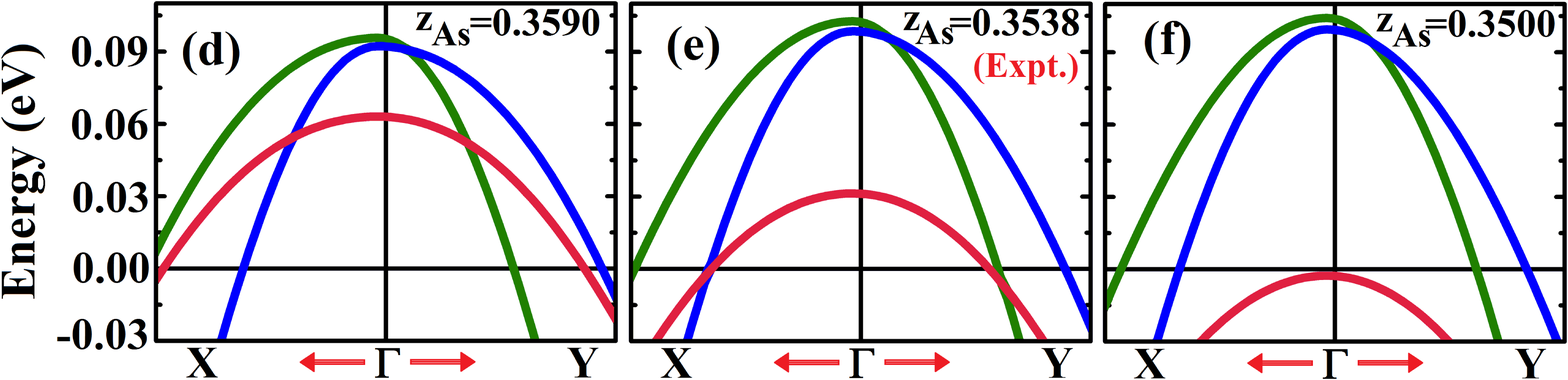}
  \caption{(Colour online) Calculated orbitally resolved zoomed BSs of  BaFe$_{2-x}$Ru$_x$As$_2$  in the low temperature orthorhombic phase for three different temperatures (top row). Calculated BSs of BaFe$_2$As$_2$ for three different values of z$_{As}$ around $\Gamma$ point.
  }
  \label{RuBS}
 \end{figure}
 
 \begin{figure}[h]
\centering
\includegraphics[width=0.5\textwidth,height=5.75cm]{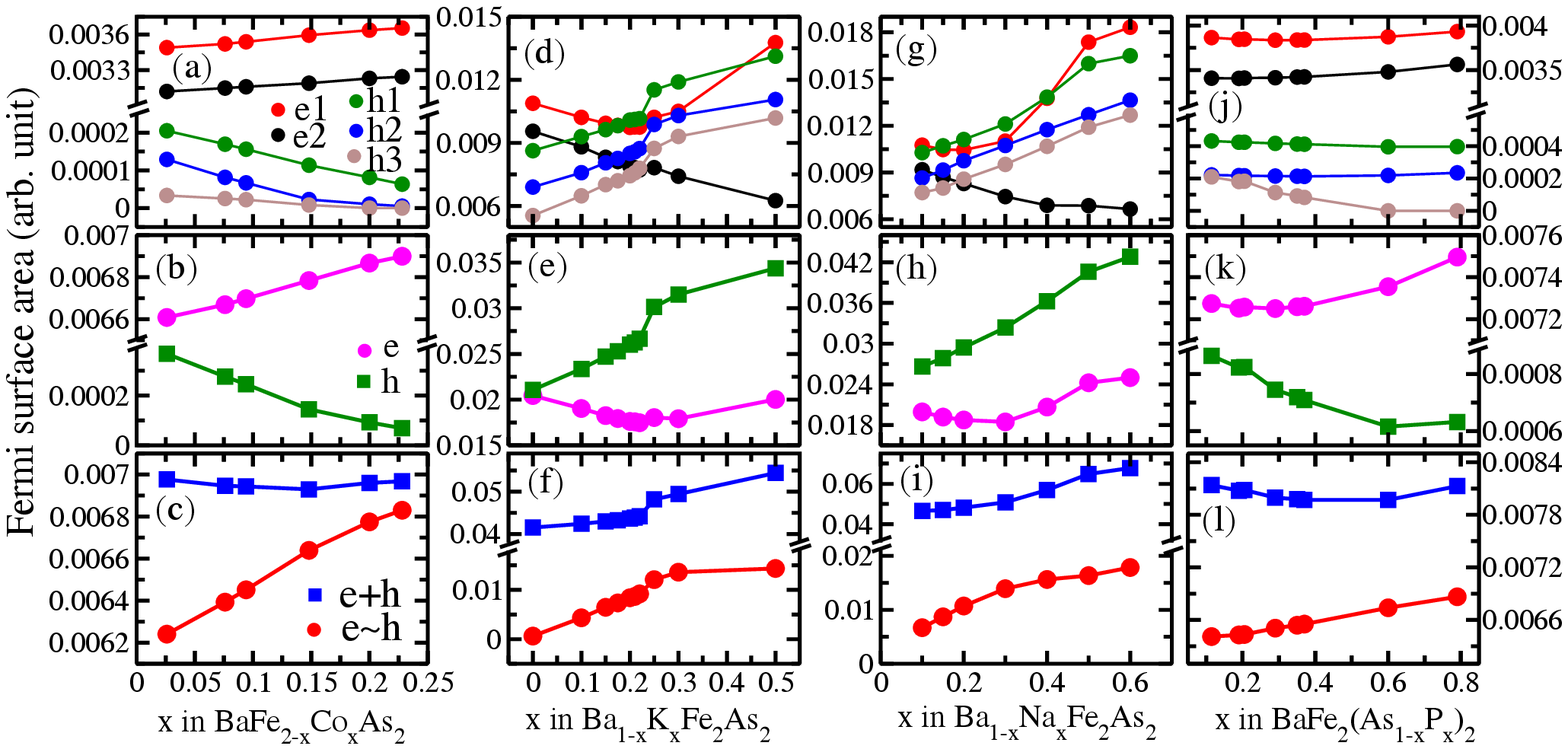}
\caption{(Colour online) FSAs of various doped Fe-based SCs 
namely Co doped (1st column), K doped (2nd column), Na doped (3rd column) and P doped (4th column), 
as a function of doping concentration. First row, FSA of each individual FSs
with doping; sudden change in slopes at the Lifshitz point is worth noting. Second row, total electron and 
hole FSAs as a function of doping; jumps/change in slopes at the Lifshitz doping is noticeable.
Third row, total FSAs and the difference in the FSAs of electron and hole FSs.
Remarkable modification of the later at the Lifshitz doping is seen.}
\label{FSA}
\end{figure}
In Fig.\ref{KFS}(d \--- f) we 
present the FSs of Ba$_{1-x}$Na$_x$Fe$_2$As$_2$ for 
various  doping concentrations ($x$). Similar to the ETT found in Ba$_{1-x}$K$_x$Fe$_2$As$_2$ materials, Ba$_{1-x}$Na$_x$Fe$_2$As$_2$ also exhibits 
the same at around the same $x$. 
BSs of Na doped Ba122 systems presented in Fig.\ref{BS-e-h-iso} (c,d) are
qualitatively very similar to that of the K doped Ba122 system. 
So we conclude from these two hole doped superconductors that the ETT are found in the electron like bands around the X and Y points at nearly 50\% hole doping. Remarkably, the established phase diagrams of the hole doped Ba122 SCs indicate that the highest  achievable T$_c$ is also around 50\% K (Na) doping where we found LT. 

In Fig.\ref{KFS} (g\---i), we display FSs of BaFe$_{2-x}$Co$_x$As$_2$ (electron doped Fe based SC) in the tetragonal phase. It is quite evident from the right column figures \ref{KFS} (g\---i) that one of the inner hole like FSs around $\Gamma$ point is modified remarkably due to Co doping. For $x=0.228$ ({\it i.e}, around 11 $\%$ of Co doping) in BaFe$_{2-x}$Co$_x$As$_2$ system, the FS around $\Gamma$ point vanishes which corresponds to a topological change in the FS, bears the signature of LT. These findings are quite consistent with experimental observation \cite{LiuA}. In Fig.\ref{BS-e-h-iso} (e\---f), we show electronic BS of BaFe$_{2-x}$Co$_x$As$_2$. The hole like (d$_{xy}$) band that lies well above FL for $x=0.026$ (Fig. \ref{BS-e-h-iso}e) moves below the FL with higher Co doping $x=0.228$ (Fig. \ref{BS-e-h-iso}f) causing doping induced topological change in the FS. Note BaFe$_{2-x}$Co$_x$As$_2$ SCs have maximum T$_c$ at around 11$ \%$ doping where LT is found from our calculation. Therefore, both the hole doped as well as electron doped systems exhibit LT where SC attains maximum T$_c$ and magnetism vanishes in Fe-based SCs. It  should also be noted here that on the contrary to hole 
doped Ba122 systems, LT in the electron doped Fe-based SCs occurs in the hole FS around $\Gamma$ point. 
Our findings are very much consistent with the previous experimental findings \cite{Liu}. 


We study the FSs and BSs of BaFe$_2$(As$_{1-x}$P$_x$)$_2$ for various isovalent P dopings in two different 
structural phases at different temperatures (low temperature orthorhombic phase and high temperature tetragonal phase). 
It is quite evident from the right column of Figs.\ref{KFS} (j\---l) that the topology of one of the hole like FSs is getting modified with increasing P doping. Our calculated orbital resolved BSs shown in Figs.\ref{BS-e-h-iso} (g\---l) distinctly reflect signatures of LT as a result of larger P doping. One of the bands, namely the d$_{xy}$ band around $\Gamma$ point, well above the FL moves below the FL. 
We compare the signatures of LT/ETT between two structural phases through orbitally resolved BSs of BaFe$_2$(As$_{1-x}$P$_x$)$_2$ in the low temperature orthorhombic phase (Fig.\ref{BS-e-h-iso}(g,h,i)) and the high temperature tetragonal phase (Fig.\ref{BS-e-h-iso} (j, k, l)) for various $x$. This part of the work thus demonstrates combined doping and temperature induced LT. 
In both the phases (10K and 300K), d$_{xy}$ band around $\Gamma$ point goes below FL at $x=0.6$. But a closer look at the BSs of BaFe$_2$(As$_{1-x}$P$_x$)$_2$ 
with $x=0.37$ at 10K and at 300K, reveal 
that there are important differences in the BSs. In case of BS of orthorhombic phase (10K) (which is in SC phase), the d$_{xy}$ 
band is about to cross the FL (Lifshitz point) but on the other hand in case of non-SC
tetragonal phase (300K) d$_{xy}$ lies above
the FL (FSAs in the two cases are different, see below). This provides an indication of the influence of temperature and structural phase, on ETT/LT. {\em Note} in this case also LT occurs at a doping where SC is maximum and no magnetism. 
From Figs.\ref{BS-e-h-iso}(g\---l) one can see that the d$_{xz}$ and d$_{yz}$ bands are hardly affected by P doping at As sites, but d$_{xy}$ bands are very sensitive to doping concentration. One of the possible reasons of this fact may arise from the structural modification caused by P doping on As site (Fe-As hybridization modifies anion height characterized by z$_{As}$). This fact is adequately demonstrated in Fig.\ref{RuBS}.


In case of BaFe$_{2-x}$Ru$_x$As$_2$ also, we find LT (not shown here for brevity) at relatively higher Ru doping 
concentration (\textgreater 50\%) where 
SC T$_c$ reaches its maximum. 
This is consistent with the findings of earlier theoretical work on Sr122 system \cite{Wang}. Below we present the temperature  induced LT in  5\% Ru doped BaFe$_2$As$_2$. 
It is well established that the electronic structure of Fe-based SCs is highly sensitive 
to temperature \cite{Sen,Dhaka,Acta,sust}. It should also be noted that the structural parameters which essentially control the electronic structures of these systems, have significant temperature dependencies \cite{Acta,Avci,Avci2,Allred}. We found temperature dependent LT in 5\% Ru doped Ba122 
system (BaFe$_{2-x}$Ru$_x$As$_2$ with $x=0.1$)  ({\it cf} Fig.\ref{RuFS}) where inner
hole FS at 90 K nearly vanishes. 
In Fig.\ref{RuBS}, we depict the BSs of BaFe$_{2-x}$Ru$_x$As$_2$  ($x=0.1$) around $\Gamma$ point 
which clearly show that the d$_{xy}$ band near FL goes below FL as temperature is lowered from 120K to 90K. Temperature induced LT is not very common in Fe-based superconductors. 
To find the root cause of temperature dependent LT, we carry out further calculations on undoped 
Ba122 \cite{Rotter} with various z$_{As}$ around the experimental value.
Essentially we find from Figs.\ref{RuFS} (d\---f),\ref{RuBS} (d\---f) that 
electronic structures very near to the FL are heavily influenced by z$_{As}$ and it reproduces the overall essential features of temperature induced LT in Ru as well as P doped Ba122 SCs. 

 FSAs as a function of various kinds of doping shown in Fig.\ref{FSA} can be experimentally measured and turns out to be 
 a unique way of identifying ETT/LT. Normally, hole doping would enlarge (reduce) hole (electron) FSA in contrast to electron doping 
  that would enlarge (reduce) electron (hole) FSA. But in case of any particular FS undergoing TT 
  it would follow against the above and such signatures are remarkably visible in  Figs.\ref{FSA}(d),(e),(g),(h),(k)). 
  The red curves in Fig.\ref{FSA} (d,g) suggest electron-FSA initially gets reduced with hole doping but increases after certain hole doping because that FS undergoes topological modification from electron like to hole like.
  This has caused sudden increase in the FSAs of hole-FSs see {\it e.g}, green, blue, pink curves of Fig.\ref{FSA} (d,e) around the LT. Thus FSA are found to carry sensitivity of topological modifications more acutely than the BSs and can be used as a better experimental tool to identify ETT/LT. The difference in FSAs of hole and electron FSs indicate net electronic charge density which increases with doping ({\it cf} red curves in Fig.\ref{FSA} (c,f,i,l)) as expected, but the slope of the curves are different for different materials. This is consistent with experiments \cite{Ideta}. The doping induced net charge density vary differently below and above the LT, much slowly above the LT indicating less effective doping above LT ({\it cf} Fig.\ref{FSA} (i,f,l))); absence of a particular FS due to LT further reduces density of states at FL and thereby high T$_c$ would be limited \cite{setti}.
 
 We have systematically shown in a variety of high T$_c$ Fe-based SCs that SC occurs 
 at the verge of LT/ETT where magnetism disappears. This is achieved through detailed evaluation of FSs through first principles simulations with experimental structural parameters as a function of doping (temperature) as inputs and detailed demonstrations of LT/ETT. The doping induced net charge density is found to be suppressed at the LT/ETT. This definitely indicates the intriguing heed to 
  the inter-relationship between SC and LT in Fe-based SCs. We have also provided a new way of detecting LT/ETT, through evaluation of FSA as a function of doping or temperature as the case may be which can be experimentally performed and is applicable in general to any system. This may modify the experimental phase diagram of Fe-SCs as far as the location of ETT/LT are concerned. We believe our work will open up many theoretical as well as experimental research activities in this direction.
 
  {\bf Acknowledgements} We thank Dr. P. A. Naik and Dr. P. D. Gupta for their encouragement in this work. One of us (SS) acknowledges the HBNI, RRCAT for financial support and encouragements.


\begin{thebibliography}{9}
\bibitem{setti} A clear case of this may be recognized from the fact that the T$_c$ of (CaFeAs)$_{10}$Pt$_{3.58}$As$_8$ is twice as high as that of
(CaFe$_{0.95}$Pt$_{0.05}$As)$_{10}$Pt$_3$As$_8$; S. Thirupathaiah  {\it et al.}, Phys. Rev. B. {\bf 88}140505 (2013).
\end{thebibliography}
\end{document}